\documentclass[aps,pre,reprint,superscriptaddress,10pt,longbibliography]{revtex4-1}
\usepackage[colorlinks,bookmarks=false,citecolor=blue,linkcolor=blue,urlcolor=blue]{hyperref}
\usepackage{amsmath, amssymb, graphicx, color}
\bibpunct{[}{]}{,}{n}{}{} %% in-line citations

\newcommand{\Tr}{\mathop{\rm Tr}\nolimits}

\begin{document}

\title{Footprints of impurity quantum phase transitions in quantum Monte Carlo statistics}

\author{Vladislav Pokorn\'y}
\email{pokornyv@fzu.cz}
\affiliation{Institute of Physics, Czech Academy of Sciences, 
Na Slovance 2, CZ-18221 Praha 8, Czech Republic}

\author{Tom\'a\v{s} Novotn\'y}
\email{tno@karlov.mff.cuni.cz}
\affiliation{Department of Condensed Matter Physics, Faculty of Mathematics and Physics, 
Charles University, Ke Karlovu 5, CZ-12116  Praha 2, Czech Republic}

\date{\today}

\begin{abstract}
Interacting single-level quantum dot connected to BCS superconducting leads
represents a well-controllable system to study the interplay between the correlation effects 
and the electron pairing that can result in a $0-\pi$ (singlet-doublet) 
quantum phase transition. The physics of this system can be well described by the
impurity Anderson model. We present an analysis of the statistics of the perturbation 
expansion order of the continuous-time hybridization expansion quantum Monte Carlo algorithm
in the vicinity of such a first-order impurity quantum phase transition. 
By calculating the moments of the histograms of the expansion order which deviate 
from the ideal Gaussian shape, we provide an insight into the thermodynamic mixing 
of the two phases at low but finite temperatures which is reflected in the bimodal 
nature of the histograms.
\end{abstract}

%\keywords{}

\maketitle

%%%%%%%%%%%%%%%%%%%%%%%%%%%%%%%%%%%%%%%%%%%%%%%%%%%%%%%%%%%
%%%%%%%%%%%%%%%%%%%%%%%%%%%%%%%%%%%%%%%%%%%%%%%%%%%%%%%%%%%
\section{Introduction \label{Sec:Intro}}
Advances in the fabrication of nanodevices have enabled us
to connect quantum dots with superconducting leads, 
forming superconducting quantum dot nanostructures. 
Many experimental realizations utilizing carbon nanotubes, 
semiconducting nanowires or even single molecules as central 
active elements demonstrate the large versatility of the concept
which attracted a lot of attention
(for the overview see Refs.~\cite{Wernsdorfer-2010,Rodero-2011,Meden-2019}).

In many cases the system can be well described by the 
single-impurity Anderson model (SIAM) coupled to 
superconducting BCS leads~\cite{Luitz-2012}. 
This model may exhibit, depending on the parameters, 
a so-called $0-\pi$ transition signaled by the sign 
reversal of the supercurrent. This transition is governed 
by the underlying quantum critical point (QCP) and can be 
induced by changing the experimentally controllable model 
parameters such as the local energy level (gate voltage) or 
superconducting phase difference (magnetic flux in SQUID geometry) 
as well as by changing parameters that are hard to control 
in experiment, e.g. the local on-dot Coulomb interaction 
or the tunneling rate between a lead and a dot.

The usual method of choice for solving the superconducting 
SIAM is the numerical renormalization group (NRG) 
method~\cite{Bulla-2008}. It works at zero and low temperatures 
and provides direct access to the one-particle spectral function. 
The drawback of this method is the inability to solve this model 
using reasonable computational resources for more than two lead 
channels, e.g., in a case with an additional metallic lead. 
Despite the recent progress in this area~\cite{Zalom-2021} it 
still hinders the usability of NRG in multi-terminal and multi-dot setups.

Another class of numerically exact methods to solve SIAM 
are the quantum Monte Carlo (QMC) techniques. Many flavours of 
QMC were already employed to solve the superconducting 
SIAM including the Hirsch-Fye~\cite{Siano-2004} 
and the continuous-time implementations in either interaction 
expansion (CT-INT)~\cite{Luitz-2010,Luitz-2012}
or hybridization expansion (CT-HYB)~\cite{Domanski-2017,Pokorny-2018} 
formalism. All these algorithms work in the imaginary time domain and are 
therefore bound to finite temperatures with unpleasant scaling properties 
with decreasing temperature. Also, obtaining the spectral function 
requires to perform an analytic continuation to real frequency domain 
from stochastic imaginary time data which is a notoriously
ill-defined problem~\cite{Jarrell-1996}. Nevertheless, QMC
can provide unbiased finite-temperature results and invaluable 
insight into the thermal effects.

Some of the most interesting effects in superconducting quantum dots are 
realized in the vicinity of the $0-\pi$ impurity quantum phase 
transition (QPT). This transition is of the first order and is 
induced by a crossing of the two lowest-lying many-body
eigenstates - a spin-singlet and a spin-doublet~\cite{Maurand-2012}.
A convenient tool to study a QPT is the so-called fidelity which 
is a measure of the overlap of ground states of quantum systems 
sharing the same Hamiltonian but associated with different Hamiltonian 
parameters~\cite{Zanardi-2006,Gu-2010,Albuquerque-2010,Rossini-2018}.
This quantity is by definition very sensitive to the change of the 
ground state induced by the variation of a control parameter. 
A method how to extract fidelity and its second derivative w.r.t the 
change of the control parameter - the fidelity susceptibility, 
from various QMC methods was used to study QPT in bulk 
(in Hubbard and Heisenberg models~\cite{Wang-2015prx,Huang-2016}) 
as well as impurity (in the two-orbital Anderson 
model~\cite{Wang-2015prl}) systems. Unfortunately the calculation 
of the fidelity susceptibility is not implemented by default
in most of the publicly available QMC solvers 
(except the iQIST package~\cite{Huang-iQIST}).

In this paper we show that a different and valuable insight into 
the properties of a system close to a QPT can be obtained by studying 
the statistics of the perturbation expansion order from a CT-HYB 
calculation. We utilize the CT-HYB method to study the $0-\pi$ 
transition in superconducting impurity Anderson model (SCIAM). 
This scenario is rather different from QPT in bulk models 
(Hubbard or Heisenberg) as we are dealing with an impurity QPT 
in a zero-dimensional system where the usual concepts 
like correlation length and finite-size scaling are not defined.

We quantify the unusual non-Gaussian shape of the histograms 
of the expansion order using the first four moments: mean, 
variance, skewness and excess kurtosis and discuss their behavior 
by changing the control parameters and the temperature. 
We show how the position of the QPT at zero temperatures can 
be extracted from the average expansion order and how is the 
thermodynamic mixing of the two phases reflected in the higher moments.
Finally, we illustrate the applicability 
of this method in the Kondo regime by providing results on the
BCS-Kondo crossover which leaves a distinct trace on the variance
of the expansion order.

The paper is organized as follows. In Sec.~\ref{Sec:Model} we 
introduce the SCIAM and the transformation that allows us to 
use CT-HYB to solve it. Sec.~\ref{Sec:Moments} introduces the moments
of the expansion order and explains how they describe the statistics 
of a random variable. In Sec.~\ref{Sec:Results} we present and discuss 
results on the behavior of the system using the local energy as 
a control parameter and their connection to previous results 
from Ref.~\cite{Kadlecova-2019}. We explain the unusual bimodal 
shape of the histogram of the expansion order and how is it reflected
in the moments. We also briefly discuss how the BCS-Kondo
crossover in the narrow-gap limit is reflected on the second statistical moment.
The main points are then
summarized in Sec.~\ref{Sec:Concl}. 
In Appendix.~\ref{App:Poisson} we discuss an alternative fit of the
histograms far from the QPT. Finally, Appendix.~\ref{App:UdepPhidep} 
presents more data obtained by changing other control parameters, 
namely the interaction strength and the superconducting phase 
difference, showing the universality of the results from 
Sec.~\ref{Sec:Results}.

%%%%%%%%%%%%%%%%%%%%%%%%%%%%%%%%%%%%%%%%%%%%%%%%%%%%%%%%%%%
%%%%%%%%%%%%%%%%%%%%%%%%%%%%%%%%%%%%%%%%%%%%%%%%%%%%%%%%%%%
\section{Superconducting impurity Anderson model \label{Sec:Model}}
\subsection{Model formulation \label{SSec:Ham}}
The Hamiltonian of a spinful, single-level quantum dot 
connected to left (L) and right (R) phase-biased superconducting 
leads reads
\begin{equation}
\label{Eq:Ham}
\mathcal{H}=\mathcal{H}_{\mathrm{dot}}+
\mathcal{H}_{\mathrm{lead}}+\mathcal{H}_{\mathrm{hyb}}
\end{equation}
where the quantum dot is described by
\begin{equation}
\label{Eq:HDot}
\mathcal{H}_{\mathrm{dot}}=\sum_\sigma \varepsilon_\sigma 
d^\dag_\sigma d^{\phantom{\dag}}_\sigma
+U \left(d^\dag_\uparrow d^{\phantom{\dag}}_\uparrow-\frac{1}{2}\right)
\left(d^\dag_\downarrow d^{\phantom{\dag}}_\downarrow-\frac{1}{2}\right) 
\end{equation}
where $d^\dag_\sigma$ creates an on-dot electron with spin 
$\sigma\in\{\uparrow,\downarrow\}$, $\varepsilon_\sigma=\varepsilon-\sigma B$ 
is the local energy level, $B$ is the magnetic field, $U$ is the on-site 
Coulomb interaction and the local energy is shifted in a way that 
$\varepsilon=0$ always represents a half-filled dot. The non-interacting 
electrons in the leads are described by BCS Hamiltonians
\begin{equation}
\label{Eq:HLead}
\begin{aligned}
\mathcal{H}_{\mathrm{lead}}&=
\sum_{\alpha\mathbf{k}\sigma}\varepsilon^{\phantom{\dag}}_{\alpha\mathbf{k}\sigma}
c^\dag_{\alpha\mathbf{k}\sigma}c^{\phantom{\dag}}_{\alpha\mathbf{k}\sigma}\\
&-\Delta\sum_{\alpha\mathbf{k}}\left(e^{i\varphi_\alpha}c^\dag_{\alpha\mathbf{k}\uparrow}
c^\dag_{\alpha\mathbf{\bar{k}}\downarrow}+\mathrm{H.c.}\right)
\end{aligned}
\end{equation}
where $c^\dag_{\alpha\mathbf{k}\sigma}$ creates an electron 
in lead $\alpha\in\{L,R\}$ with spin $\sigma$ and energy 
$\varepsilon_{\alpha\mathbf{k}\sigma}$, and 
$\Delta e^{i\varphi_\alpha}$ is the complex superconducting order 
parameter for lead $\alpha$ with amplitude $\Delta$ and phase 
$\varphi_\alpha$. Here we assume that the amplitude is the same 
in both leads (i.e. they are made from the same material)
but the phases can differ. We note that physical observables 
can only depend on the phase difference $\varphi=\varphi_L-\varphi_R$ 
and not on the individual values, i.e. the system must be invariant 
w.r.t. a global phase shift 
$\varphi_\alpha\rightarrow\varphi_\alpha+\varphi_{sh}$~\cite{Meden-2019}.
We also denoted $\mathbf{\bar{k}}=-\mathbf{k}$ to save space in equations.
The coupling between the dot and the leads is governed by
\begin{equation}
\label{Eq:HHyb}
\mathcal{H}_{\mathrm{hyb}}=-\sum_{\alpha\mathbf{k}\sigma}
\left(V^{\phantom{\dag}}_{\alpha\mathbf{k}\sigma}
c^\dag_{\alpha\mathbf{k}\sigma}d^{\phantom{\dag}}_\sigma+\mathrm{H.c.}\right)
\end{equation}
where $V_{\alpha\mathbf{k}\sigma}$ is the tunnel matrix element.

As the Hamiltonian in~\eqref{Eq:Ham} does not conserve charge, we 
cannot use the standard CT-HYB algorithm right away to solve it. 
Therefore we perform a canonical particle-hole transformation in 
the spin-down segment of the Hilbert space. Following 
Ref.~\cite{Luitz-2010} we define
\begin{equation}
\begin{aligned}
\left(\tilde{c}^{\phantom{\dag}}_{\alpha\mathbf{k}\uparrow},
\tilde{c}^\dag_{\alpha\mathbf{k}\uparrow},
\tilde{c}^{\phantom{\dag}}_{\alpha\mathbf{k}\downarrow},
\tilde{c}^\dag_{\alpha\mathbf{k}\downarrow}\right)&=
\left(c^{\phantom{\dag}}_{\alpha\mathbf{k}\uparrow},
c^\dag_{\alpha\mathbf{k}\uparrow},
c^\dag_{\alpha\mathbf{\bar{k}}\downarrow},
c^{\phantom{\dag}}_{\alpha\mathbf{\bar{k}}\downarrow}\right),\\
\left(\tilde{d}^{\phantom{\dag}}_\uparrow,\tilde{d}^\dag_\uparrow,
\tilde{d}^{\phantom{\dag}}_\downarrow,\tilde{d}^\dag_\downarrow\right)&=
\left(d^{\phantom{\dag}}_\uparrow,d^\dag_\uparrow,
d^\dag_\downarrow,d^{\phantom{\dag}}_\downarrow\right)
\end{aligned}
\end{equation}
which maps the SCIAM~\eqref{Eq:Ham} on the SIAM with
non-superconducting bath but with attractive on-site 
interaction strength $-U$. We can define Nambu-like spinors in this basis as 
\begin{equation}
\tilde{D}=
\begin{pmatrix}
\tilde{d}_\uparrow \\[0.3em]
\tilde{d}_\downarrow
\end{pmatrix},
\quad
\tilde{C}_{\alpha\mathbf{k}}=\begin{pmatrix}
\tilde{c}_{\alpha\mathbf{k}\uparrow} \\[0.3em]
\tilde{c}_{\alpha\mathbf{\bar{k}}\downarrow}
\end{pmatrix}.
\end{equation}
From now on we drop the spin index in the dispersion $\varepsilon_{\alpha\mathbf{k}}$ 
and the tunnel coupling element $V_{\alpha\mathbf{k}}$. Let us define matrices
\begin{equation}
E_{\alpha\mathbf{k}}=
\begin{pmatrix}
\varepsilon_{\alpha\mathbf{k}} & -\Delta e^{i\varphi_\alpha} \\[0.3em]
-\Delta e^{-i\varphi_\alpha} & -\varepsilon_{\alpha\mathbf{\bar{k}}}
\end{pmatrix},\quad
E_d=
\begin{pmatrix}
\varepsilon_\uparrow & 0 \\[0.3em]
0 & U-\varepsilon_\downarrow
\end{pmatrix}.
\end{equation}
The Hamiltonian in~\eqref{Eq:Ham} can be rewritten in the form
\begin{equation}
\begin{aligned}
\mathcal{H}_{\mathrm{dot}}&=\tilde{D}^\dag E_d \tilde{D}
-U \left(\tilde{d}^\dag_\uparrow \tilde{d}^{\phantom{\dag}}_\uparrow-\frac{1}{2}\right)
\left(\tilde{d}^\dag_\downarrow \tilde{d}^{\phantom{\dag}}_\downarrow+\frac{1}{2}\right), \\
\mathcal{H}_{\mathrm{lead}}&=\sum_{\alpha\mathbf{k}}
\tilde{C}^\dag_{\alpha\mathbf{k}}E_{\alpha\mathbf{k}} \tilde{C}_{\alpha\mathbf{k}},\\
\mathcal{H}_{\mathrm{hyb}}&=-\sum_{\alpha\mathbf{k}}
\left(\tilde{C}^\dag_{\alpha\mathbf{k}}V_{\alpha\mathbf{k}}\sigma_z\tilde{D}+\mathrm{H.c.}\right)
\end{aligned}
\end{equation}
where $\sigma_z$ is a Pauli matrix.
The occupation numbers transform as
\begin{equation}
\label{Eq:densities}
\begin{aligned}
n&=n_\uparrow+n_\downarrow=\tilde{n}_\uparrow-\tilde{n}_\downarrow+1 \\
m&=n_\uparrow-n_\downarrow=\tilde{n}_\uparrow+\tilde{n}_\downarrow-1 \\
\nu&=\tilde{\nu} 
\end{aligned}
\end{equation}
where $n_\sigma=\langle d^\dag_\sigma d^{\phantom{\dag}}_\sigma\rangle$ 
is the spin-resolved electron density, $n$ is the total charge, 
$m$ is the magnetization and $\nu=\langle d_\downarrow d_\uparrow\rangle$ 
is the on-dot induced pairing.

From now on we assume the tunneling density of states (DOS) to be constant 
in the relevant energy window, 
\begin{equation}\label{Eq:TDOS}
A_{T\alpha}(\omega)=
\pi\sum_{\mathbf{k}}|V_{\alpha\mathbf{k}}|^2\delta(\omega-\varepsilon_{\alpha\mathbf{k}})
=\Gamma_\alpha\Theta(W^2-\omega^2)
\end{equation} 
where $W$ is the half-bandwidth of the non-interacting band. 
As the relevant energy scale for this model is the 
superconducting gap $\Delta$ which is usually of order of 
$100~\mu$eV, the assumption of constant DOS over such a narrow 
energy window is justifiable in most experimentally relevant 
situations.
Equation~\eqref{Eq:TDOS}
also defines the tunneling rates $\Gamma_\alpha$ which, for 
$\mathbf{k}$-independent tunnel matrix elements $V$ can be 
expressed as $\Gamma_\alpha=\pi|V_{\alpha}|^2/(2W)$.
We will focus only on the situations with symmetric coupling 
$\Gamma_L=\Gamma_R=\Gamma/2$ as any asymmetric setup with 
$\Gamma_L\neq\Gamma_R$ can be mapped on the symmetric one 
using a simple formula~\cite{Kadlecova-2017}. 

The total effect of a lead $\alpha$ on the impurity can be 
summed into a hybridization function which describes the hopping 
from the impurity to the lead $\alpha$, the propagation through the
lead and the hopping back to the impurity. It can be represented 
in the imaginary-frequency (Matsubara) domain as
\begin{equation}
\Gamma_{\alpha}(i\omega_n)=
\sum_\mathbf{k} V_{\alpha\mathbf{k}}^*\sigma_z
[i\omega_nI_2-E_{\alpha\mathbf{k}}]^{-1}\sigma_zV_{\alpha\mathbf{k}}
\end{equation}
where $I_2$ is the $2\times 2$ unit matrix, 
$\omega_n=(2n+1)\pi/\beta$ is the $n$th Matsubara frequency 
and $\beta=1/(k_BT)$ is the inverse temperature. 
Under the assumption of a constant tunneling DOS~\eqref{Eq:TDOS}, 
the hybridization function of the lead $\alpha\in\{L,R\}$ can be expressed as
\begin{equation}
\Gamma_{\alpha}(i\omega_n)=
-\frac{\Gamma_\alpha w(i\omega_n)}{\sqrt{\omega_n^2+\Delta^2}}
\begin{pmatrix}
i\omega_n & \Delta e^{i\phi_\alpha} \\[0.3em]
\Delta e^{-i\phi_\alpha} & i\omega_n
\end{pmatrix}
\end{equation}
where $w(i\omega_n)=(2/\pi)\arctan(W/\sqrt{\omega_n^2+\Delta^2})$ 
is the correction due to finite bandwidth that approaches 
unity for $W\rightarrow\infty$.

\subsection{$0-\pi$ quantum phase transition \label{SSec:QPT}}
The so-called $0-\pi$ QPT corresponds to the change 
of the ground state from a non-magnetic singlet ($0$ phase) 
to a spin-degenerate doublet ($\pi$ phase). 
It is signaled at zero temperatures by the 
abrupt change of the sign of both the Josephson current and the on-dot 
induced pairing from positive in $0$-phase to negative 
in $\pi$ phase~\cite{vanDam-2006,Cleuziou-2006,Jorgensen-2007},
and the crossing of the subgap, Andreev bound states at the 
Fermi energy~\cite{Pillet-2010,Pillet-2013,Li-2017}. As this transition
is of the first order, the finite-temperature state is a simple 
thermodynamic mixture of the two phases and the transition is smeared
out by increasing temperature into a crossover.

The method of extraction of the position of the QPT 
at zero temperature from finite temperature (experimental or QMC) 
data was already discussed in Ref.~\cite{Kadlecova-2019}. 
For low temperatures, only the lowest-lying many-body states are 
significant. Due to the presence of a superconducting pairing, 
these states are discrete and separated from the continuous part 
by a gap of the order of $\Delta$. In the absence of magnetic field, 
we are dealing with a one spin singlet and one spin doublet. 
The second discrete singlet state and the continuum of states
lie higher in energy and can be neglected at low temperatures. 
The canonical average of an observable $F$ then reads
\begin{equation}
\begin{aligned}
\label{Eq:avg1}
\langle F(x)\rangle&=\frac{1}{\mathcal{Z}}\sum_n F_n(x)e^{-\beta E_n(x)} \\
&\approx\frac{F_s(x)e^{-\beta E_s(x)}
+2F_d(x)e^{-\beta E_d(x)}}{e^{-\beta E_s(x)}+2e^{-\beta E_d(x)}}
\end{aligned}
\end{equation}
where $\mathcal{Z}$ is the partition function,
$F_s$ ($F_d$) are zero-temperature values of $F$ in singlet 
(doublet) state and $x$ is a control parameter 
(e.g. local energy $\varepsilon$), changing of which crosses the QPT 
at critical value $x_c$. At QPT, the two discrete many-body states cross, 
i.e., $E_s(x_c)=E_d(x_c)$ and the average of an observable $F$ reads
\begin{equation}
\label{Eq:avg2}
\langle F(x_c)\rangle=\frac{1}{3}\left[F_s(x_c)+2F_d(x_c)\right].
\end{equation}
This value is independent of temperature, which implies that all 
curves $F(x)$ for low enough temperatures cross at one point and this 
point marks the position of the QPT at zero temperatures.

%%%%%%%%%%%%%%%%%%%%%%%%%%%%%%%%%%%%%%%%%%%%%%%%%%%%%%%%%%%
%%%%%%%%%%%%%%%%%%%%%%%%%%%%%%%%%%%%%%%%%%%%%%%%%%%%%%%%%%%
\section{Moments of the expansion order \label{Sec:Moments}}
A complete overview of the CT-HYB method can be found 
in Ref~\cite{Gull-2011}. It is based on an expansion of the 
partition function in powers of the hybridization term 
$\mathcal{H}_{\mathrm{hyb}}$~\eqref{Eq:HHyb} around 
$\mathcal{H}_{\mathrm{dot}}+\mathcal{H}_{\mathrm{lead}}$.
The order of the expansion $k$ (the number of hybridization 
events in the given random configuration) depends rather strongly 
on the model parameters. It scales linearly with inverse temperature 
$\beta$ and in general decreases with increasing interaction 
strength $U$. The statistics of $k$ can be accumulated during the 
Monte Carlo run and represented in the form of a histogram $H(k)$. 
The moments of $k$ can be then calculated and their analysis provides 
a deeper insight into the performance of the solver and also the 
properties of the solved model.

We focus on four moments of the expansion order: 
mean, variance, skewness and excess kurtosis. 
The behavior of the variance and the excess kurtosis of $k$ 
around a phase transition point has been briefly discussed by 
Huang \textit{et al.} in an appendix of 
Ref.~\cite{Huang-2016} for the case of the Mott transition in the single-band 
Hubbard model within the dynamical mean-field theory (DMFT). 
As the $0-\pi$ impurity QPT in the superconducting Anderson model is a way 
simpler scenario than a Mott transition in the Hubbard model, it provides
an opportunity to examine their behavior in a well-controllable setup
and shed more light on the statistics of the expansion order.

The $n$th central moment of a discrete random variable $k\in\{0,1,2\ldots\}$ 
is defined as
\begin{equation}
\label{Eq:moments}
\mu_n\equiv\langle (k-c_n)^n\rangle=\frac{1}{N}\sum_{k=0}^\infty (k-c_n)^n p(k)
\end{equation}
%where $p(k)/N$ is the probability density function,
where $p(k)$ is the number of occurrences of given $k$ in the sample,
$c_n$ is the $n$th central value and $N$ is the number of measurements.
The mean $\mu=\mu_1$ is the first moment about zero ($c_1=0$),
and variance as the second moment about the mean ($c_2=\mu$), 
\begin{equation}
\label{Eq:variance}
\sigma^2=\langle (k-\mu)^2\rangle=\langle k^2\rangle-\mu^2.
\end{equation}
For higher moments it is useful to define \textit{standardized moments} 
(normalized moments about the mean) $\bar{\mu}_n=\mu_n/\sigma^n$. 
Skewness and kurtosis are then defined as the third and fourth standardized moments, 
$\gamma=\bar{\mu}_3$ and $\kappa_0=\bar{\mu}_4$, respectively.
As the kurtosis of the Gaussian distribution equals three, it is often 
convenient to define the excess kurtosis $\kappa=\kappa_0-3$. To summarize,
\begin{equation}
\label{Eq:moments2}
\begin{aligned}
\mu&=\langle k\rangle,\qquad\qquad\quad
\sigma^2=\langle (k-\mu)^2\rangle,\\
\gamma&=\frac{\langle (k-\mu)^3\rangle}{\sigma^3},\qquad
\kappa=\frac{\langle (k-\mu)^4\rangle}{\sigma^4}-3.
\end{aligned}
\end{equation}

\begin{figure}
\includegraphics[width=1.0\columnwidth]{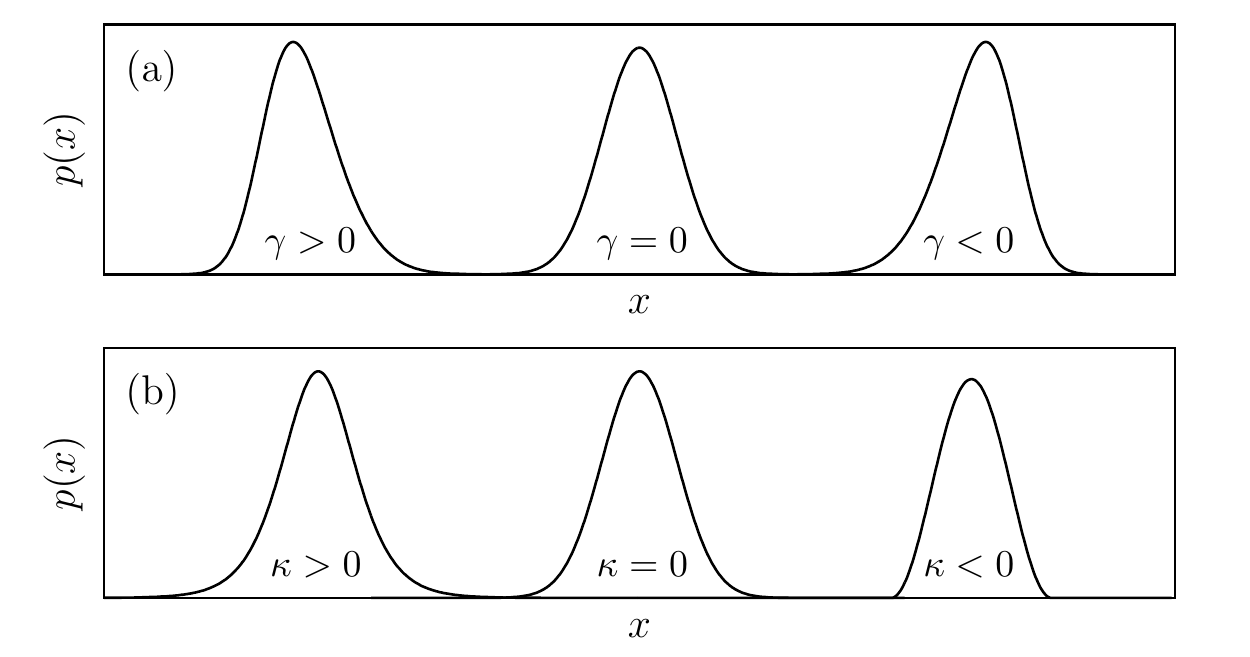}
\caption{Examples of distribution functions 
$p(x)$ with different skewness $\gamma$ (panel a) 
and excess kurtosis $\kappa$ (panel b).
\label{Fig:examples}}
\end{figure}

The first two moments are standard tools in mathematical 
statistics and their meaning is clear, but skewness and 
kurtosis are more abstract quantities. Therefore we illustrate 
them in Fig.~\ref{Fig:examples}. Skewness measures the asymmetry 
of the tails of a distribution. In panel (a) we plotted distributions 
with skewness $\gamma>0$ (left-leaning, right tail is longer), 
$\gamma=0$ (no skew) and $\gamma<0$ (right-leaning, left tail is longer). 
The excess kurtosis measures how fast the distribution vanishes 
at higher values compared to the Gaussian. Distributions with 
$\kappa>0$ (with ``fat tails'') are usually referred to as leptokurtic,
with $\kappa=0$ as mesokurtic and with $\kappa<0$ (``thin tails'') 
as platykurtic. Examples of such distributions are plotted in panel (b).

It is worth noting that an alternative way to describe a 
distribution, complementary to the moments, is with statistical 
cumulants. The first four cumulants $K_n$ of a random variable 
$k$ are connected to the standardized moments as defined in 
Eq.~\eqref{Eq:moments2} as 
$K_1=\mu$, $K_2=\sigma^2$, $K_3=\gamma\sigma^3$, and $K_4=\kappa\sigma^4$,
and the cumulant analysis would be analogous to the moment 
analysis presented below.

The averaged expansion order $\mu$ from CT-HYB also has a physical meaning. 
In the case of the impurity Anderson model it is connected to the 
hybridization energy as 
$\langle k \rangle=\beta E_{\mathrm{hyb}}$ where~\cite{Haule-2007}
\begin{equation}
\label{Eq:Ehyb}
E_{\mathrm{hyb}}=\langle\mathcal{H}_{\mathrm{hyb}}\rangle\equiv
\frac{1}{\mathcal{Z}}\Tr\left[e^{-\beta\mathcal{H}}\mathcal{H}_{\mathrm{hyb}}\right].
\end{equation}
This is one of the differences between a calculation 
of the impurity Anderson model and a DMFT calculation of 
the Hubbard model, for which the averaged expansion order 
from CT-HYB can be identified with the kinetic energy. 
Here, $E_{\mathrm{hyb}}$ can also be expressed in terms of the 
local interacting Green function 
$G(i\omega_n)=[i\omega_nI_2-E_{d}-\Sigma(i\omega_n)]^{-1}$ 
and the hybridization function $\Gamma(i\omega_n)$. 
Here $\Sigma(i\omega_n)$ is the dynamical self-energy induced 
by $U$. From the equation of motion we obtain~\cite{Merker-2012}
\begin{equation}
E_{\mathrm{hyb}}=-\frac{1}{\beta}\Tr\sum_{n}\Gamma(i\omega_n)G(i\omega_n).
\end{equation}
This formula can be evaluated numerically and used as a 
benchmark of the accuracy of the measurement of the 
local Green function w.r.t. the averaged expansion order.

%%%%%%%%%%%%%%%%%%%%%%%%%%%%%
%%%%%%%%%%%%%%%%%%%%%%%%%%%%%

%%%%%%%%%%%%%%%%%%%%%%%%%%%%%%%%%%%%%%%%%%%%%%%%%%%%%%%%%%%
%%%%%%%%%%%%%%%%%%%%%%%%%%%%%%%%%%%%%%%%%%%%%%%%%%%%%%%%%%%
\section{Results \label{Sec:Results}}
All CT-HYB calculations were performed using the TRIQS/CTHYB 2.2 solver~\cite{Seth-2016}. 
We set $B=0$, $W=100\Delta$ and the cutoff in Matsubara frequencies 
$\omega_n^{\mathrm{max}}\geq 300\Delta$.
We encountered no fermionic sign problem during the calculations. 
The total charge $n$ and the induced pairing $\nu$
were evaluated by tracing the measured impurity density matrix 
and using Eq.~\eqref{Eq:densities}. 

%%%%%%%%%%%%%%%%%%%%%%%%%%%%%%%%%%%%%%%%%%%%%%%%%%%%%%%%%%%
\subsection{Shape of histograms around the QPT \label{SSec:Moments}}
\begin{figure}
\includegraphics[width=1.0\columnwidth]{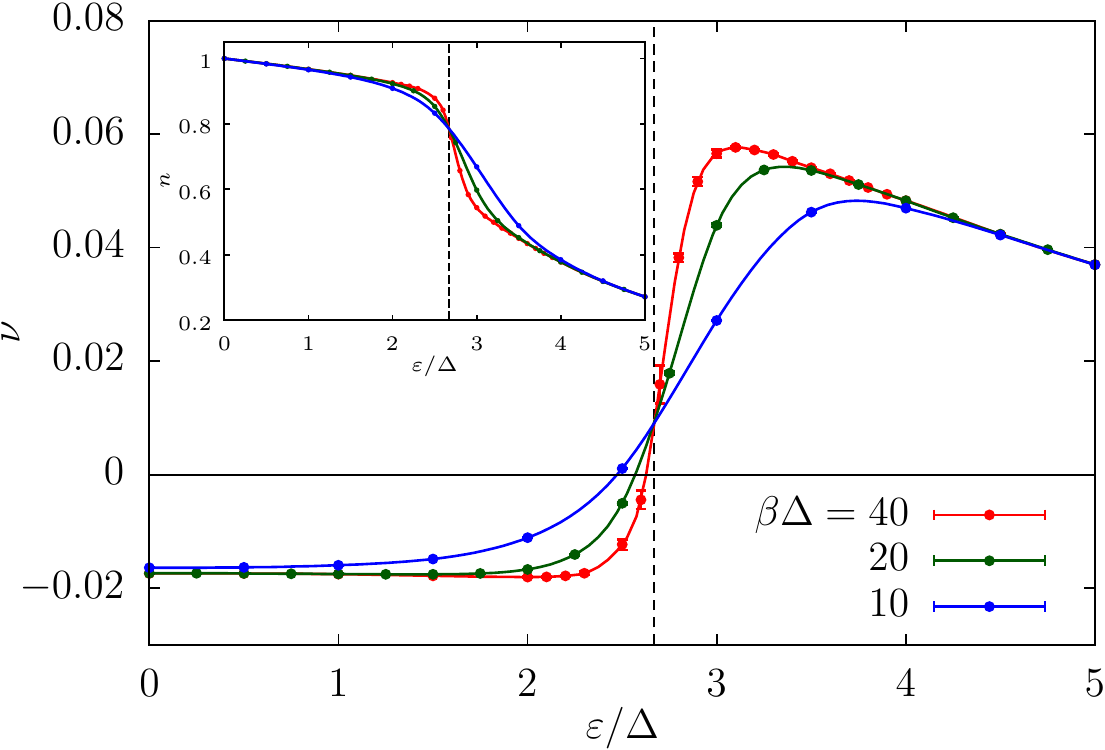}
\caption{On-dot induced pairing $\nu$ as a function of the 
local energy $\varepsilon$ calculated using CT-HYB for $U=8\Delta$, $\Gamma=\Delta$ 
and $\varphi=0$ for three values of inverse temperature 
$\beta\Delta=40$, $20$ and $10$. 
Lines are splines of QMC data and serve only as a guide for the eye.
All curves cross at $\varepsilon_c\approx2.67\Delta$ within the error bars.
Inset: On-dot electron density $n$ for the same set of parameters, showing 
the same crossing behavior around the critical point as the induced pairing.
\label{Fig:mu_eps}}
\end{figure}

In Fig.~\ref{Fig:mu_eps} we plotted the on-dot induced pairing 
$\nu$ as a function of the local energy $\varepsilon$
for $U=8\Delta$, $\Gamma=\Delta$ and $\varphi=0$ 
for three values of inverse temperature $\beta\Delta=40$, $20$ and $10$.
Here $\varepsilon=0$ represents the half-filled dot.
This plot illustrates the typical behavior of SCIAM around the QPT.
The induced paring is positive in the $0$-phase and negative 
in the $\pi$-phase and all lines cross at $\varepsilon_c\approx2.67\Delta$. 
The position of such a crossing is consistent with the 
position of the QPT at zero-temperature
as explained in Ref.~\cite{Kadlecova-2019}.
The inset shows the total on-dot electron density $n$ 
that decreases as we move away from the half-filled
case. Again, all lines cross at the same value of 
$\varepsilon$, proving the applicability of the
formula in~\eqref{Eq:avg2}.

%%%%%%%%%%%%%%%%%%%%%%%%%%%%%
%%%%%%%%%%%%%%%%%%%%%%%%%%%%%

\begin{figure}
\includegraphics[width=1.0\columnwidth]{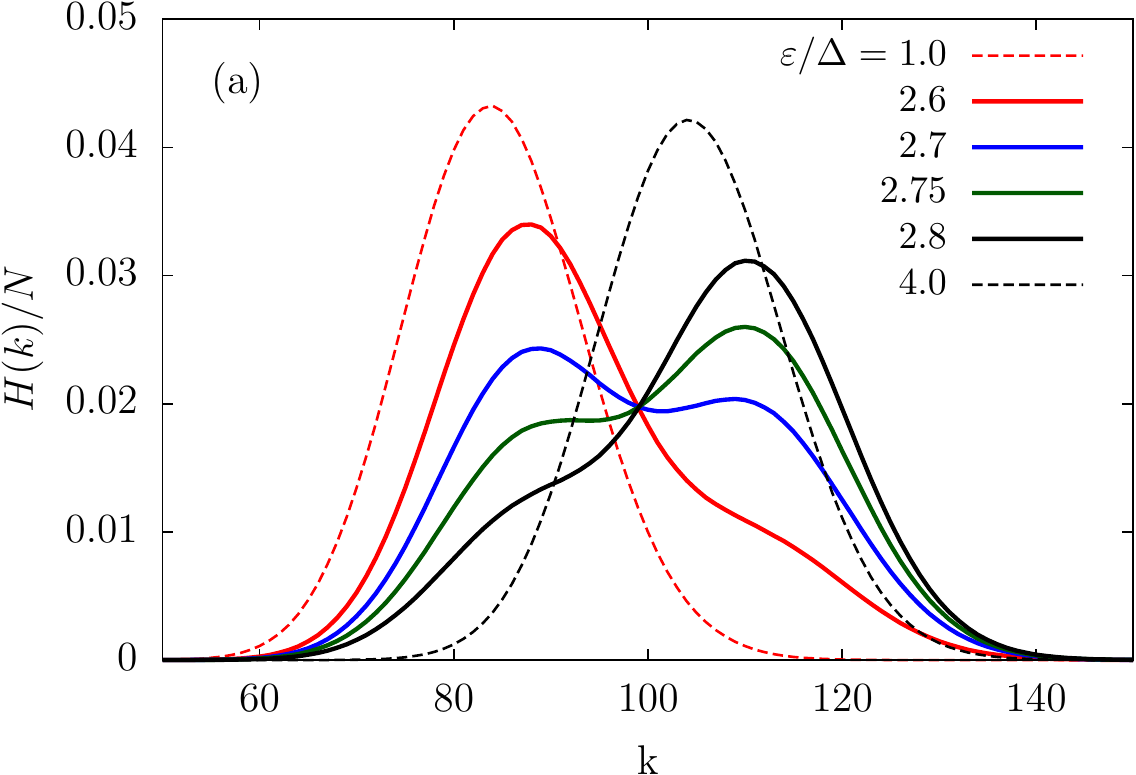}\\
\vspace{5mm}
\includegraphics[width=1.0\columnwidth]{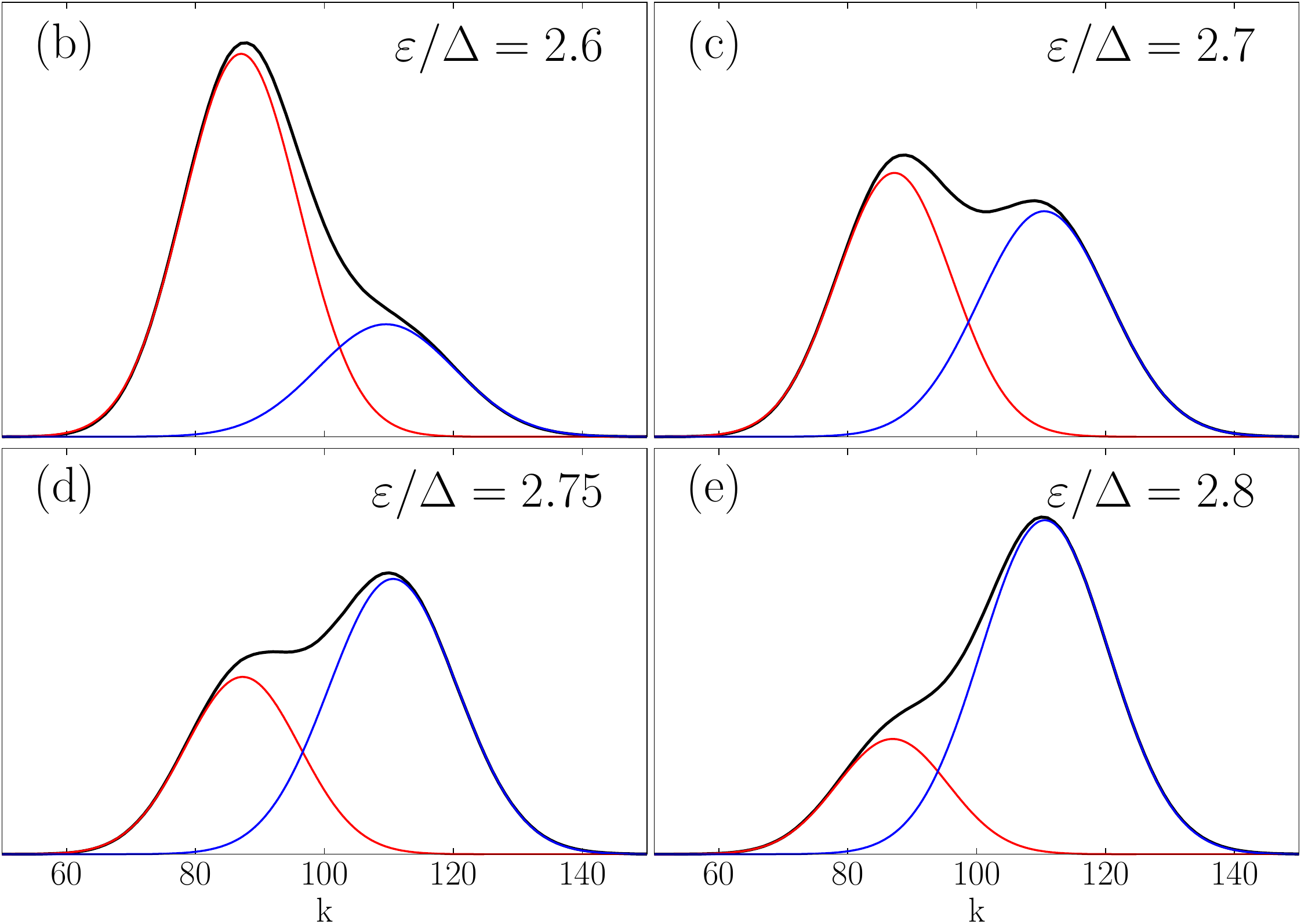}
\caption{Panel (a): Histograms of the expansion order $k$ 
from CT-HYB calculation for different values of
phase difference around the zero-temperature critical value 
$\varepsilon_c\approx 2.67\Delta$ for the same parameters
as in Fig.~\ref{Fig:mu_eps}, $U=8\Delta$, $\Gamma=\Delta$ and $\varphi=0$ 
and inverse temperature $\beta\Delta=40$. All histograms for $\varepsilon$ close to
$\varepsilon_c$ (solid lines) show a non-Gaussian, two-peak 
structure and intersect in one point for 
$k\approx 99$. We added the Gaussian-like histograms 
for the two values far from the transition point,
$\varepsilon=\Delta$ and $\varepsilon=4\Delta$ (dashed lines) for comparison.
Panels (b)-(e): Histograms from panel (a) fitted with a pair of Gaussians.
Red (blue): contribution from the doublet (singlet) ground state.
\label{Fig:histo_eps}}
\end{figure}

The histograms of the expansion order $k$ 
normalized to the number of QMC measurements ($4.8\times 10^8$ in this case) 
for $\beta\Delta=40$ are plotted in Fig.~\ref{Fig:histo_eps}(a)
We plotted few histograms around the critical value $\varepsilon_c$ 
(solid lines) together with histograms for values further from the transition point,
$\varepsilon=\Delta$ and $\varepsilon=4\Delta$ (dashed lines) for comparison. 
As the averaged expansion order is rather large, the histograms resemble
more a continuous curve than a discrete set of values.
It is also the reason why the histograms are not distorted much by 
the fact that the expansion order is bounded from below by zero.
As the averaged expansion order increases with decreasing
temperature, it is always possible to run the simulation 
at low enough temperatures to minimize the distortion.

We see that histograms for $\varepsilon=\Delta$ and $4\Delta$ are of the Gaussian shape. 
This shape does not change until we approach 
the critical point. As we move closer, the histogram starts to deviate strongly
from a Gaussian, first developing a shoulder and later turning into a broad 
two-peaked structure. Similar double-peaked histograms were also reported in
Ref.~\cite{Huang-2016} for the Hubbard model in the vicinity of the Mott transition.
The histograms for values of $\varepsilon$ close to the critical value also 
intersect at one point at $k\approx 99$.
As the two-peak structure can be well fitted by a pair of Gaussians,
we can assume these histograms can be also seen as thermal averages 
over the singlet and doublet ground states,
\begin{equation}
\label{Eq:avg_pt_histo}
\begin{aligned}
H(k,\varepsilon)\approx\frac{H_s(k,\varepsilon)e^{-\beta E_s(\varepsilon)}
+2H_d(k,\varepsilon)e^{-\beta E_d(\varepsilon)}}
{e^{-\beta E_s(\varepsilon)}+2e^{-\beta E_d(\varepsilon)}}.
\end{aligned}
\end{equation}
We denote the intersection point of the two histograms as $k_0$, so 
$H_s(k_0,\varepsilon_c)=H_d(k_0,\varepsilon_c)$ where we replaced the 
local energy by its critical value, assuming that the constituent histograms $H_s$ and $H_d$ 
do not depend much on $\varepsilon$ around the critical point.
As $E_d(\varepsilon_c)=E_s(\varepsilon_c)$, we obtain in the vicinity of 
the critical point ($\varepsilon\approx\varepsilon_c$),
\begin{equation}
H(k_0,\varepsilon)=H_s(k_0,\varepsilon)=H_d(k_0,\varepsilon)
\end{equation}
for all values of $\varepsilon$ close enough to the critical point that 
we can replace $H_{s,d}(k,\varepsilon)\approx H_{s,d}(k,\varepsilon_c)$. This feature represents 
itself as the histograms crossing at the same point at $k=k_0$ 
for $\varepsilon\approx\varepsilon_c$. In panels (b)-(e) we plotted selected histograms 
in the vicinity of the critical point fitted by a pair of Gaussians.
This illustrates the feature that the constituent histograms $H_s$ and $H_d$
do not depend much on $\varepsilon$, only their weights 
are changing as we cross the transition point.

%%%%%%%%%%%%%%%%%%%%%%%%%%%%%
%%%%%%%%%%%%%%%%%%%%%%%%%%%%%

%%%%%%%%%%%%%%%%%%%%%%%%%%%%%
\begin{figure*}
\includegraphics[width=2.0\columnwidth]{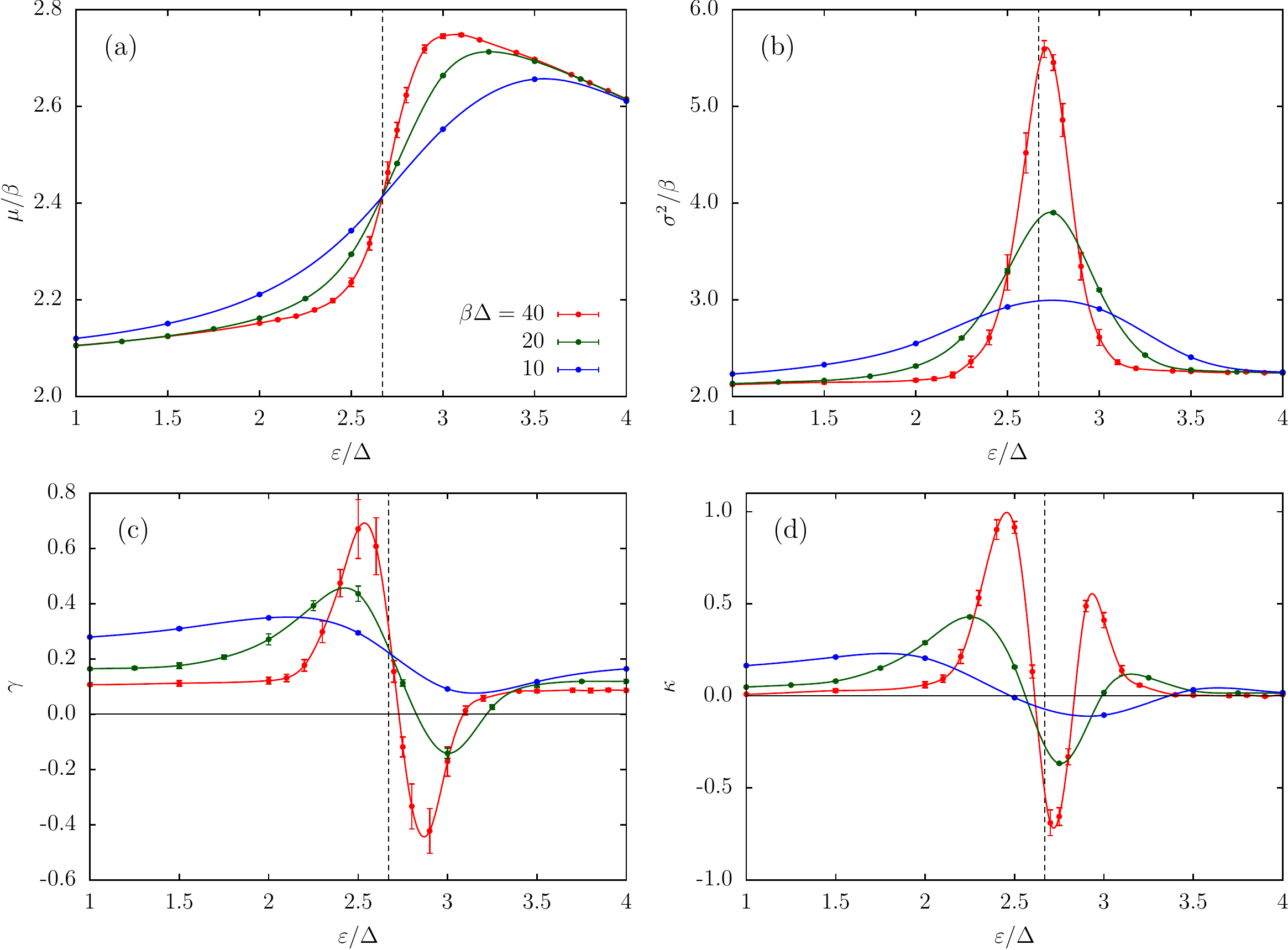}
\caption{Mean (panel a), variance (b), skewness (c) and excess kurtosis (d) 
of the expansion order $k$ calculated from the histograms provided by the 
CT-HYB solver as functions of the local energy $\varepsilon$ 
for the same parameters as in Fig.~\ref{Fig:mu_eps}, 
$U=8\Delta$, $\Gamma=\Delta$ and $\varphi=0$. 
Lines are splines of QMC data and serve only as a guide for the eye. 
\label{Fig:po_eps}}
\end{figure*}
%%%%%%%%%%%%%%%%%%%%%%%%%%%%%

The first four standardized moments (mean, variance, skewness and excess kurtosis)
as defined by Eq.~\eqref{Eq:moments2} calculated from the 
histograms for the same parameters as in Figs.~\ref{Fig:mu_eps},\ref{Fig:histo_eps}
are plotted in Fig.~\ref{Fig:po_eps}.
These quantities were calculated using Eq.~\eqref{Eq:moments} 
where we approximated $p(k)$ by the histogram $H(k)$. 
Panel (a) shows the behavior of 
the average expansion order $\mu$ scaled to inverse temperature $\beta$
for three values $\beta\Delta=40$, $20$ and $10$. All curves 
cross at $\varepsilon_c$, like the curves for the induced pairing. 
This is not a surprise since we identified the physical meaning 
of $\mu/\beta$ as the hybridization energy and at low temperatures 
Eq.~\eqref{Eq:Ehyb} also reduces to a relation in a form of 
Eq.~\eqref{Eq:avg2}. It proves that the position of the QCP is 
encoded already in the statistics of the expansion order and can 
be obtained without actually measuring any physical observable
or the one-particle Green function.

We can obtain the same result that lines $\mu(\varepsilon)/\beta$ 
for different low enough temperatures cross 
at the same point using the assumption that the histogram in the vicinity of 
the critical point can be decomposed into parts from the singlet and from the doublet, 
$H(k,\varepsilon\approx\varepsilon_c)=[H_s(k,\varepsilon)+2H_d(k,\varepsilon)]/3$. 
Then Eq.~\eqref{Eq:moments} for $n=1$ and $c_1=0$ 
reduces to a weighted sum of two contributions,
$\mu\equiv\langle k \rangle=(\langle k \rangle_s+2\langle k \rangle_d)/3$
where $\langle k \rangle_{s,d}$ is an average of $k$ w.r.t. $H_{s,d}$.

The broadening of the histogram can be quantified by 
the scaled variance $\sigma^2/\beta$ plotted in Fig.~\ref{Fig:po_eps}(b)
that shows a peak above $\varepsilon_c$. 
For $\beta\Delta=40$ the peak lies at $\varepsilon\approx2.72\Delta$ 
(between the blue and green curves in Fig.~\ref{Fig:histo_eps}(a)). 
The curves for different temperatures do not cross at the same point. 
Using the same reasoning as for the averaged expansion order we obtain
\begin{equation}
\label{Eq:var}
\begin{aligned}
\sigma^2&\approx\frac{1}{3}\Big[\langle k^2\rangle_s+2\langle k^2\rangle_d
-\frac{1}{9}\left(\langle k\rangle_s+2\langle k\rangle_d\right)^2\Big]\\
&=\frac{1}{3}(\sigma_s^2+2\sigma_d^2)+\frac{2}{9}(\langle k\rangle_s-\langle k\rangle_d)^2.
\end{aligned}
\end{equation}
The first term scales as $\beta$ with temperature while 
the second scales as $\beta^2$. As a result, curves of scaled variance 
$\sigma^2/\beta$ at the transition point do not cross as there is
a linear offset due to the last term in Eq.~\eqref{Eq:var}. 
As the width of the crossover 
region between the two phases is increasing with increasing 
temperature, the peak in variance becomes wider and less pronounced 
and the maximum moves away from $\varepsilon_c$. 
As this maximum lies away from the transition point at any finite
temperature, variance is not a good estimator of the position of the QPT.

We would like to point out that the fact that the scaled average
expansion order $\mu/\beta$ can be identified with the 
hybridization energy $E_{\mathrm{hyb}}$
does not automatically imply that the variance of the expansion order
is identical to the variance of $E_{\mathrm{hyb}}$.
The question whether the variance has a physical meaning 
is beyond the scope of this paper.

The overall shape of the histogram can be well quantified 
by the skewness $\gamma$ which is plotted in Fig.~\ref{Fig:po_eps}(c). 
Histograms for low temperatures almost always have a slight
positive skewness (leaning to the left). As we approach the 
QPT by moving away from half-filling (from the $\pi$-phase to the $0$-phase), 
skewness shows a prominent positive peak connected with the development 
of the shoulder on the histogram indicating increasing presence 
of the $0$-phase in the thermal mixture. 
The skewness is zero at the same point where variance has 
a maximum, above the zero-temperature QPT. 
This is the point where the histogram is symmetric due to 
equal admixtures of the zero and the $\pi$ phases
at given temperature. 
As we move across this point, the scenario is reversed with a negative 
skewness indicating decreasing presence of the $\pi$ phase.
All the features are quickly smeared out by increasing temperature
and for $\beta\Delta=10$ the skewness is positive for all values of 
$\varepsilon$. 

The overall positive skewness of the histograms
for all temperatures indicates a systematic deviation from the 
ideal Gaussian shape even far away from the transition point. 
An alternative fitting of the histograms with a 
single-parametric Poisson curve that would explain this behavior
is discussed in Appendix~\ref{App:Poisson}.

The curves for different temperatures seem to cross at 
one point very close to $\varepsilon_c$ within the QMC error bars. 
As skewness in our definition is normalized to $\sigma^3$, an analysis 
analogous to the case of the variance in \eqref{Eq:var} becomes very 
tedious and effectively not feasible.However, simple analytical models 
of appropriate Boltzmann mixtures of two Gaussian or Poisson distributions 
exhibit qualitatively identical behavior with an apparent common crossing 
very close to the QPT. Detailed analysis reveals that it is not an exact 
single crossing, but just series of nearby ones, whose overall position 
depends on a subtle interplay of cumulants of constituent distributions 
with the Boltzmann factors. As such, this phenomenon appears to have a 
rather generic origin and does not reveal any new information about the 
microscopic features of the underlying model.

The excess kurtosis $\kappa$ provides additional 
insight into the deviations from a Gaussian shape.
It is, in contrast to the skewness, insensitive to which 
tail of the distribution deviates from the Gaussian, 
therefore the curve is rather symmetric around $\varepsilon_c$.
Excess kurtosis is almost zero at low temperatures and far 
from the transition point indicating that the tails of the 
histogram are of Gaussian shape. As we move closer to QPT 
from any side, the histogram becomes leptokurtic (fat-tailed), 
again because of the presence of the shoulder, as evident
from Figs.~\ref{Fig:histo_eps}(b),(e). This indicates that 
configurations with a large or small number of hybridization events 
are encountered here more often during the QMC sampling.
Then it shows a well developed minimum at the same point 
where skewness crosses zero, i.e. where the histogram is symmetric. 

In summary, the abovementioned statistical moments quantify well
the shape of the histogram of the expansion order. It changes its shape
from nearly ideal Gaussian to bimodal which can be fitted by a pair of
Gaussians as a result of the thermodynamic mixing of the two phases
at low but finite temperatures. The position of the zero-temperature
QCP can be extracted from the crossing of the curves of the first moment
for two low-enough temperatures. The extremal points of higher moments 
(maximum of $\sigma^2$, zero of $\gamma$, minimum of $\kappa$)
always lie away from the transition point in the direction 
of the $0$-phase and approach it only slowly with decreasing
temperature, therefore they are not good estimators of the position 
of the QCP, although they provide a way to quantify the mixing of 
the phases in the crossover region.

%%%%%%%%%%%%%%%%%%%%%%%%%%%%%%%%%%%%%%%%%%%%%%%%%%%%%%%%%%%
\subsection{BCS-Kondo crossover \label{SSec:BCS-Kondo}}
The moments analysis presented above is largely based on the two-level 
approximation from Ref.~\cite{Kadlecova-2019} which requires the temperature
to be much less than the superconducting gap $\Delta$ so that the continuous parts
of the spectra can be neglected in the close vicinity of the QPT. 
This hinders the effectiveness of this method
in the strong-coupling Kondo regime where a narrow gap is required, otherwise the 
complete Kondo screening is impossible due to the lack of the spectral weight
around the Fermi energy~\cite{Meden-2019}.

\begin{figure}
\includegraphics[width=1.0\columnwidth]{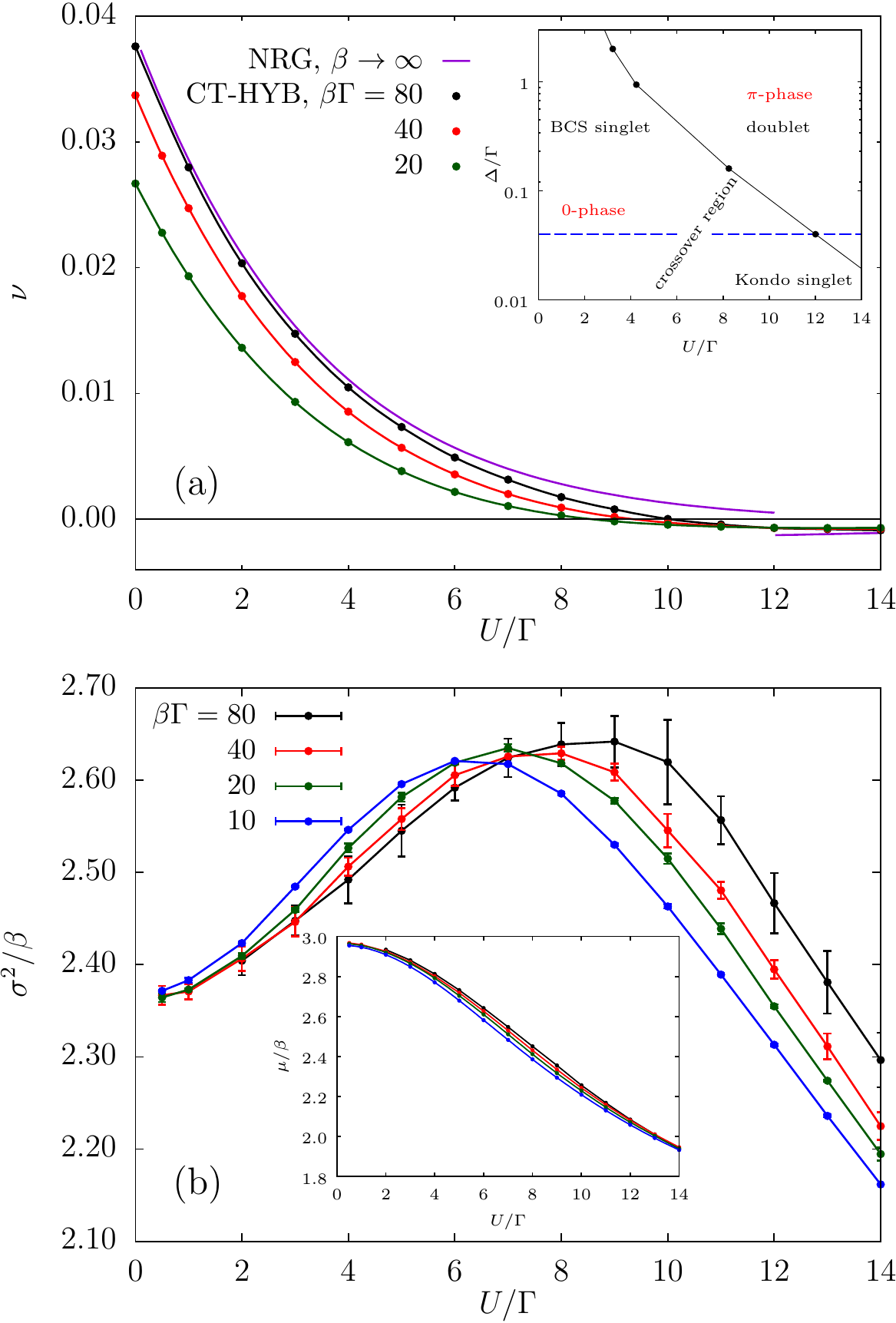}
\caption{Panel (a): Induced pairing $\nu$ as a function of interaction 
strength $U/\Gamma$ in the narrow-gap limit $\Delta/\Gamma=0.04$, 
$\varphi=0$ and $\varepsilon=0$ (half-filling) calculated using CT-HYB 
for three temperatures and compared with the zero-temperature NRG data
taken from Ref.~\cite{Zonda-2016}. QMC error bars are smaller than the 
symbol size. 
Inset: Phase diagram in the $U-\Delta$ plane. Black bullets are the $0-\pi$
phase boundary at zero temperature calculated using NRG.
Blue dashed line marks the cut along which the data in the main plot are 
calculated. Note the logarithmic scale on the vertical axis.
Panel (b): scaled variance $\sigma^2/\beta$ of the perturbation order $k$
for the same parameters as the induced gap in panel (a) that shows a broad 
peak in the crossover region between the BCS-like and the Kondo-like 
singlet regions. 
Inset: Scaled mean $\mu/\beta$ for the same parameters as in the main plot.
Lines connecting the QMC data points serve only as guides for the eye.
\label{Fig:delta_cut}}
\end{figure}

To study the limits of the usefulness of our method in the Kondo regime,
we calculated the behavior of the induced gap and the first two moments of 
the expansion order at half-filling 
in a narrow-gap limit $\Delta/\Gamma=0.04$ for $\varphi=0$.
The results are plotted in Fig.~\ref{Fig:delta_cut}. 

The inset of panel (a) shows the 
zero-temperature phase diagram in the $U-\Delta$ plane
calculated using NRG taken from Ref.~\cite{Zonda-2016}.
The nature of the ground state in the singlet
phase is governed by the competition between the superconductivity and the Kondo effect
and this phase can be in general separated into a region of large gap and small interaction 
strength where the ground state is 'BCS-like' and the 'Kondo-like' region 
of small gap size and large interaction strength.
These two regions are separated by a broad crossover region which is hard to locate
as it leaves no trace on the quantities like the electron filling or the induced gap
and can be identified usually only from the change of the 
shape of the spectral function~\cite{Yoshioka-2000}.

Panel (a) shows the induced pairing $\nu$ calculated along the blue dashed line in the inset
for three inverse temperatures 
$\beta\Gamma=20$, $40$ and $80$ compared with the zero-temperature NRG result.
The $0-\pi$ QPT lies at $U_c\approx 12.02\Gamma$. The curves for the two
lowest temperatures cross at the transition point while the curve for $\beta\Gamma=20$
is slightly off as this temperature is too high for the applicability of the two-level
approximation. The other feature, the crossover between the two types of 
singlet ground state that takes place at $U\sim6-8\Gamma$, leaves no trace on 
the induced pairing.

The scaled variance $\sigma^2/\beta$ plotted in panel (b) shows no peak
around $U_c$ that we could connect with the $0-\pi$ QPT.
As the superconducting gap $\Delta$ in this plot is 20-times smaller than in 
Figs.~\ref{Fig:mu_eps}-\ref{Fig:po_eps}, we would need to perform the simulation at 
inverse temperatures $\beta\Gamma>400$ to obtain similar
resolution around the $0-\pi$ QPT as before, which is impossible with the 
current imaginary-time implementations of the CT-HYB algorithm.
On the other hand, the variance for $\beta\Gamma=10$, $20$ and $40$
shows a broad peak in the region where the BCS-Kondo crossover is expected.
This peak moves slowly to higher interaction strengths with decreasing temperature.
The result for $\beta\Gamma=80$ shows a different profile with flatter top
and a more significant shift to higher values of interaction strengths
for $U>9\Gamma$. This can be caused by the additional effect of the 
QCP at $U_c$ that starts to have effect on the statistics 
as we move closer to it by decreasing the temperature.

We thus believe that the variance of the expansion order represents
a limited but simple tool to identify a presence of the BCS-Kondo 
crossover and provides a rough estimate of its position. While we 
cannot completely rule out the possibility that the peak in the 
variance is caused solely by the presence of the QCP at $U_c$ and 
is not connected whatsoever to the BCS-Kondo crossover, 
the different shape of the variance curve for $\beta\Gamma=80$ 
suggests there are two different phenomena having effect on the 
statistics. We would need data for much lower temperatures 
to resolve this issue which hinders the applicability
of the presented method in the narrow-gap regime. 

%%%%%%%%%%%%%%%%%%%%%%%%%%%%%
%%%%%%%%%%%%%%%%%%%%%%%%%%%%%

%%%%%%%%%%%%%%%%%%%%%%%%%%%%%%%%%%%%%%%%%%%%%%%%%%%%%%%%%%%
%%%%%%%%%%%%%%%%%%%%%%%%%%%%%%%%%%%%%%%%%%%%%%%%%%%%%%%%%%%
\section{Conclusions\label{Sec:Concl}}
Superconducting quantum dots provide a simple testbed for 
studying impurity QPT and this physics can be captured very well 
by the SCIAM. Numerically exact CT-HYB QMC simulation 
can provide an unbiased insight into finite-temperature behavior in 
the vicinity of the critical point. 
The change of the ground state from non-magnetic singlet to 
spin-degenerate doublet leaves a unique footprint on the statistics 
of the hybridization expansion order which is accumulated during the 
simulation and can be quantified by the moments of its histogram.

The position of the zero-temperature QCP can be extracted from 
finite-temperature results by measuring two sets of data for any 
physical observable at two different low-enough temperatures as 
already discussed in Ref.~\cite{Kadlecova-2019}. 
As the average expansion order in CT-HYB can be identified with 
the hybridization energy, it is also a physical observable and 
therefore position of the QCP can be extracted from its statistics 
without performing any QMC measurement at all (except a trivial 
measurement of the average sign). In the vicinity
of the critical point the histogram of the expansion order deviates 
from the ideal Gaussian shape and this deviation can be quantified by the higher 
moments (variance, skewness, excess kurtosis). We show that the 
histogram can be separated into Gaussian contributions from the 
two ground states. The weights of these contributions equalize at 
the point that linearly approaches the QCP with decreasing temperature and 
can be identified from the maximum of variance, zero of the 
skewness or the minimum of the excess kurtosis.
While this method is due to limitations on the 
lowest achievable temperature practically applicable 
only in situations with relatively large superconducting gap,
limited information about the nature of the ground
state can be extracted also in the narrow-gap limit.

We believe this analysis can provide additional insight into the 
physics around an impurity QPT beyond the scope of standard tools 
like the fidelity susceptibility for systems that can be mapped 
onto an impurity Anderson model. Furthermore, this analysis can be 
performed using any available CT-HYB solver without any modification 
or implementation of new features. The results also hold for any
system exhibiting a first-order impurity QPT
as well as bulk systems that are treated using the DMFT technique of
mapping a lattice model on an effective impurity problem.

\begin{acknowledgments}
We thank M. \v{Z}onda for valuable discussions and for providing NRG data.
This work was supported by Grant  No. 19-13525S of the Czech Science Foundation (T.N.), 
by the Ministry of Education, Youth and Sports program INTER-COST, Grant No. LTC19045 (V.P.),
by the COST Action NANOCOHYBRI (CA16218) (T.N.),
the National Science Centre (NCN, Poland) via Grant No.\ UMO-2017/27/B/ST3/01911 (T.N.),
and from the Large Infrastructures for Research, Experimental Development and Innovations 
project ``IT4Innovations National Supercomputing Center - LM2015070'' 
and  project  ``e-Infrastruktura  CZ'' - e-INFRA LM2018140.
\end{acknowledgments}

\appendix

\section{Alternative fit of the histograms\label{App:Poisson}}
\begin{figure}
\includegraphics[width=1.0\columnwidth]{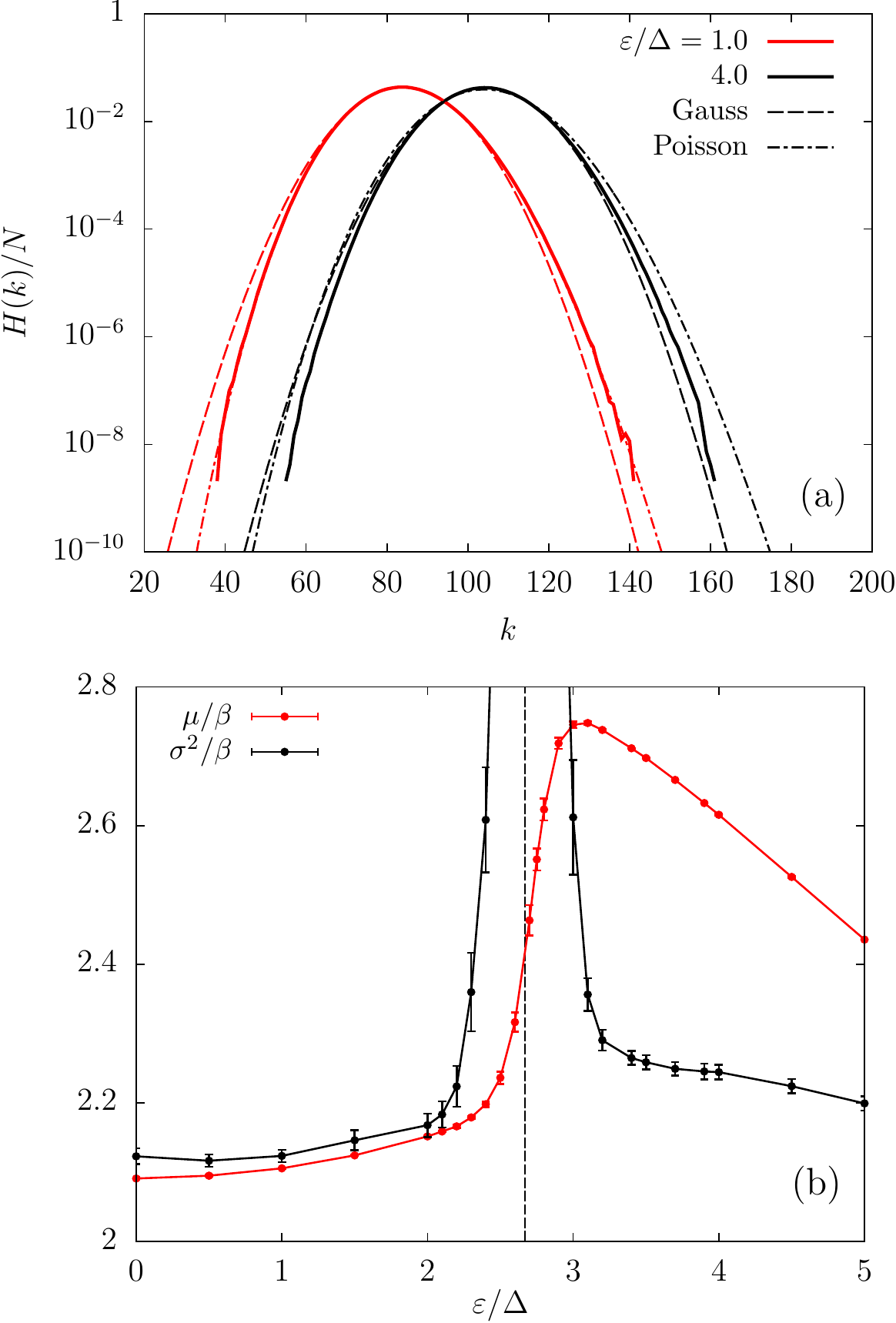}
\caption{Panel (a): Histograms far from the QPT 
from Fig.~\ref{Fig:histo_eps}a for $\beta\Delta=40$ and
parameters $\varepsilon=\Delta$ ($\pi$-phase, red) and 
$\varepsilon=4\Delta$ ($0$-phase, black) on logarithmic scale
fitted by a two-parametric Gaussian function (dashed lines)
and by a single-parametric Poisson function (dash-dotted lines).
Panel (b): Mean $\mu/\beta$ (red bullets) and variance $\sigma^2/\beta$ (black bullets)
for $\beta\Delta=40$ from Figs.~\ref{Fig:po_eps}a and~\ref{Fig:po_eps}b 
plotted in the same figure. Lines are just a guide to the eye.
Values of the mean and the variance
in the $\pi$-phase ($\varepsilon\lesssim 2\Delta$) almost coincide, 
showing the region of validity of the single-parametric Poisson fit.
\label{Fig:histo_log}}
\end{figure}

The overall positive skewness of the histograms even far from the transition
point suggests their inherent non-Gaussian nature. To look deeper into
this feature we replotted in Fig.~\ref{Fig:histo_log}a two histograms 
from Fig.~\ref{Fig:histo_eps}a for parameters far from the QPT,
$\varepsilon=\Delta$ ($\pi$-phase, red) and $\varepsilon=4\Delta$ ($0$-phase, black), 
on a logarithmic scale together with the Gaussian fit 
$p(k;\mu,\sigma^2)=\exp[-(k-\mu)^2/(2\sigma^2)]/\sqrt{2\pi\sigma^2}$ 
(dashed lines). The deviations from the Gaussian shape in both cases are evident.

Considering that the
expansion order $k$ is a discrete variable bound from below by zero and 
assuming that the QMC configurations with a given expansion 
order are generated as independent rare events,
we tried an alternative fitting with the Poisson distribution 
$\bar{p}(k;\lambda)=\lambda^ke^{-\lambda}/k!$ (dash-dotted lines).
A distinct feature of the 
Poisson distribution compared with the Gaussian is that its mean equals the variance,
$\mu=\sigma^2=\lambda$ so the fitting is single-parametric.
As the limit of $\bar{p}(k;\lambda)$
for large $\lambda$ is Gaussian, it is problematic to distinguish the two
cases for large mean values and the main clue is the inherent
single-parametric nature of $\bar{p}(k;\lambda)$ compared with the two-parametric 
function $p(k;\mu,\sigma^2)$.

We see that the histogram for $\varepsilon=\Delta$ ($\pi$-phase)
can be very well fitted with a Poisson curve, meanwhile the histogram for
$\varepsilon=4\Delta$ ($0$-phase) cannot. To identify the region 
of validity of this alternative fit we utilize the above mentioned feature
of the Poison distribution that $\mu=\sigma^2$ and we plotted in Fig.~\ref{Fig:histo_log}b
data from Figs.~\ref{Fig:po_eps}a and~\ref{Fig:po_eps}b for $\beta\Delta=40$
into a single plot. The two quantities almost coincide for 
$0<\varepsilon\lesssim 2\Delta$, that is in the $\pi$-phase (doublet ground state).
The values in the $0$-phase are different which confirms the failure of
the alternative fit for $\varepsilon=4\Delta$  (black dash-dotted line 
in Fig.~\ref{Fig:histo_log}a). 
Resolving the issue why the statistics
of the expansion order follows a Poisson distribution in only one of the phases would 
require a detailed analysis of the configurations encountered during the simulation 
and also an independent check by a different CT-HYB solver to rule out the effect of 
the given implementation of the algorithm, which is beyond the scope of this paper.

%%%%%%%%%%%%%%%%%%%%%%%%%%%%%
%%%%%%%%%%%%%%%%%%%%%%%%%%%%%

\section{Dependence on the interaction strength 
and phase difference\label{App:UdepPhidep}}
\begin{figure*}
\vspace{3cm}
\includegraphics[width=2.0\columnwidth]{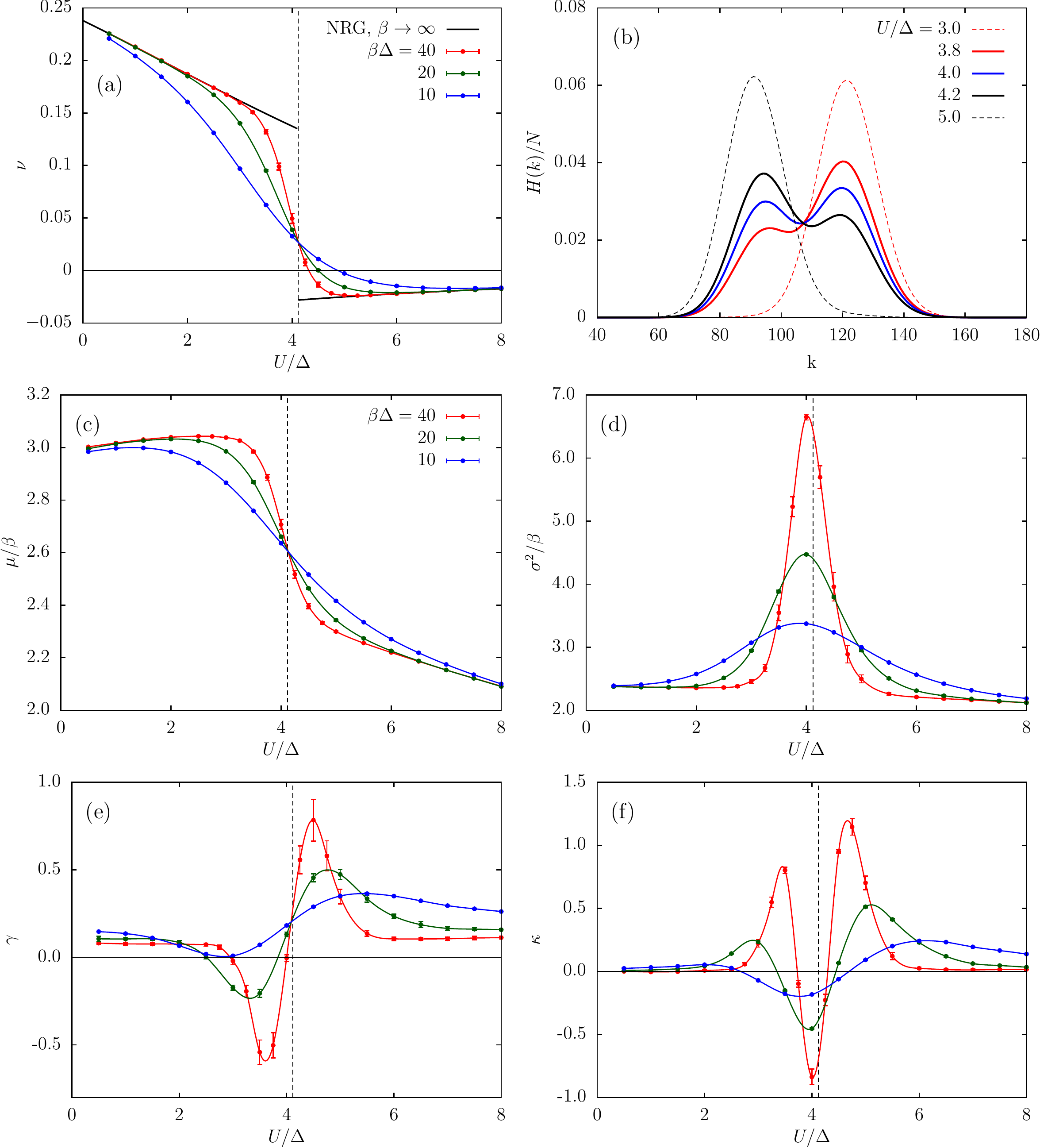}
\caption{Same analysis as in Figs.~\ref{Fig:mu_eps}-\ref{Fig:po_eps} 
but for changing interaction strength $U$ for 
$\Gamma=\Delta$, $\varphi=0$ and $\varepsilon=0$ (half-filling):
induced gap (panel a), 
histograms for $\beta\Delta=40$ in the vicinity of the critical point (b), 
average expansion order (c), 
variance (d), 
skewness (e) 
and excess kurtosis (f).
Lines are splines of QMC data and serve only as a guide for the eye.
The critical interaction strength $U_c\approx 4.12\Delta$ matches 
the available zero-temperature NRG data (solid black line in panel (a)). 
\label{Fig:po_U}}
\end{figure*}

\begin{figure*}
\vspace{3cm}
\includegraphics[width=2.0\columnwidth]{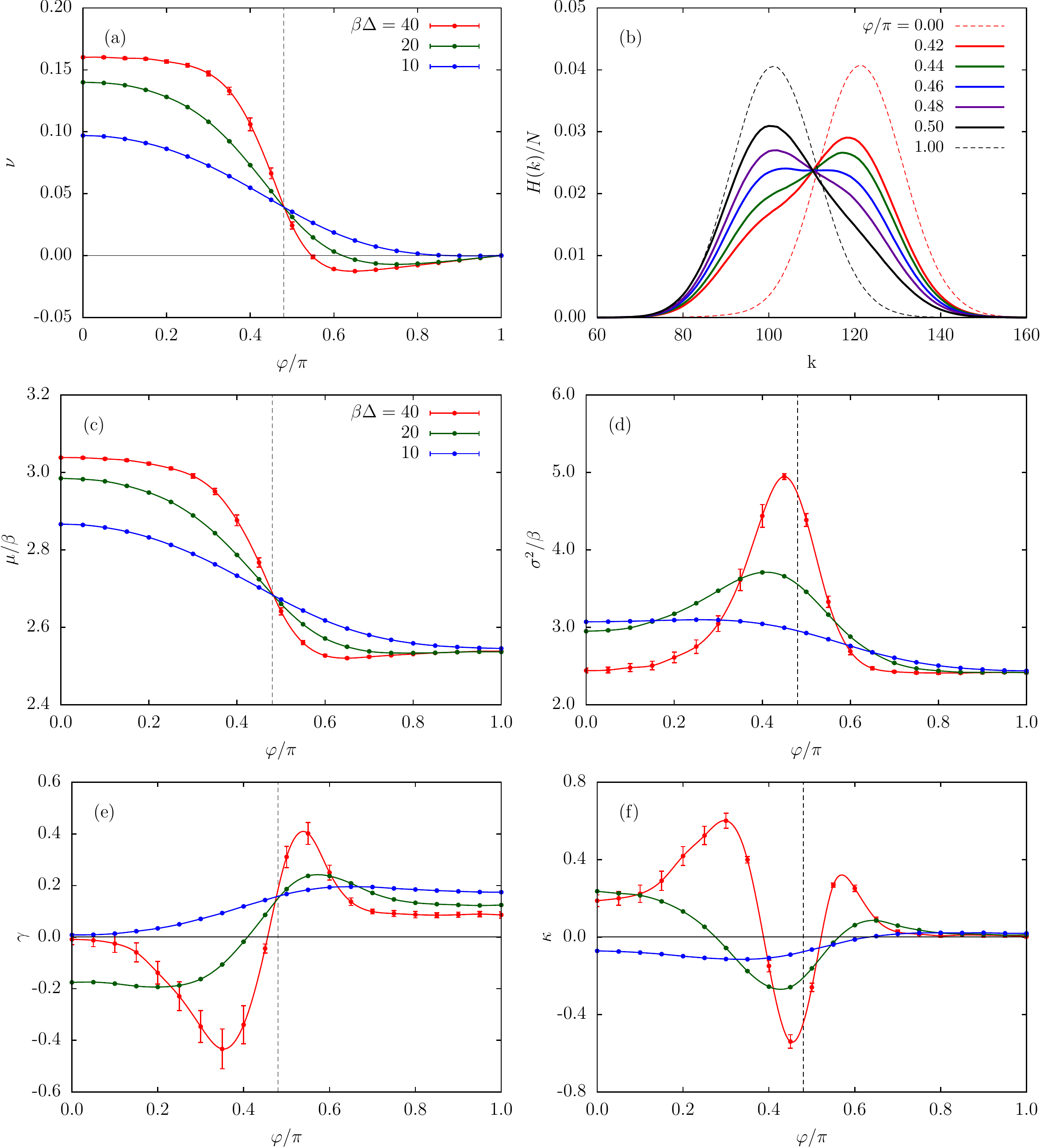}
\caption{Same analysis as in Figs.~\ref{Fig:mu_eps}-\ref{Fig:po_eps} 
but for changing phase difference $\varphi$ 
for $U=3\Delta$, $\Gamma=\Delta$ and $\varepsilon=0$ (half-filling):
induced gap (panel a), 
histograms for $\beta\Delta=40$ in the vicinity of the critical point (b), 
average expansion order (c), 
variance (d), 
skewness (e) 
and excess kurtosis (f).
Lines are splines of QMC data and serve only as a guide for the eye.
The critical phase difference $\varphi_c\approx 0.48\pi$ matches 
the available zero-temperature NRG data (not plotted).
\label{Fig:po_phi}}
\end{figure*}

A superconducting quantum dot can be driven through a QPT by 
changing several control parameters. Here we provide additional 
data on the behavior of the histograms and moments of the expansion 
order around the QPT as functions of the interaction strength 
$U$ (Fig.~\ref{Fig:po_U})
and phase difference $\varphi$ (Fig.~\ref{Fig:po_phi}). 
The behavior is analogous to the case discussed in section~\ref{SSec:Moments},
except that here we move in both cases from zero to the $\pi$-phase by 
increasing the parameter value. In both cases, panel (a) shows the 
induced pairing for three different temperatures. The curves cross 
at the same point proving the applicability of the two-level model
for the selected temperatures. We added the available zero-temperature 
NRG result to Fig.~\ref{Fig:po_U} to further illustrate this feature.
Panel (b) shows histograms for $\beta\Delta=40$ in the vicinity of the 
QPT (solid lines) that also cross at one point, supplemented by 
Gaussian-like histograms from calculations away from the QPT 
(dashed lines). Panels (c-e) show the behavior of the first four 
standardized moments as defined in Eq.~\eqref{Eq:moments2}. 
The analysis is again analogous to the $\varepsilon$-dependent 
results from the main text.

%%%%%%%%%%%%%%%%%%%%%%%%%%%%%
%%%%%%%%%%%%%%%%%%%%%%%%%%%%%

%%%%%%%%%%%%%%%%%%%%%%%%%%%%%
%%%%%%%%%%%%%%%%%%%%%%%%%%%%%

%\bibliography{pert_qmc}{}
%\bibliographystyle{apsrev4-1}

%

\end{document}